\documentclass[12pt]{amsart}
\usepackage{amssymb,latexsym}
\theoremstyle{plain}
\numberwithin{equation}{section}
\newtheorem{theorem}{Theorem}
\newtheorem{proposition}{Proposition}
\newtheorem{lemma}{Lemma}
\newtheorem{corollary}{Corollary}

\newtheorem{remark}{Remark}

\begin{document}

\title{The Quantum Black-Scholes equation} 
\author{Luigi Accardi}
\address{Centro Vito Volterra, Universit\`{a} di Roma Torvergata\\
            via Columbia  2, 00133 Roma, Italy}
\email{accardi@volterra.mat.uniroma2.it}
\urladdr{http://volterra.mat.uniroma2.it}
\author{Andreas Boukas}
\address{Department of Mathematics and Natural Sciences, American College of Greece\\
 Aghia Paraskevi, Athens 15342, Greece}
\email{andreasboukas@acgmail.gr}


\date{\today}

\maketitle

\begin{abstract}
 Motivated by the work of Segal and Segal in \cite{5} on the Black-Scholes pricing formula in the quantum context,  we study a quantum extension of the Black-Scholes equation within the context of Hudson-Parthasarathy quantum stochastic calculus,. Our model includes stock markets described by quantum Brownian motion and Poisson process.
\end{abstract}

\section{The Merton-Black-Scholes Option Pricing Model }

 An \textit{option} is a ticket which is bought at time $t=0$ and which allows the buyer at (in the case of \textit{European call} options) or until (in the case of \textit{American call} options) time $t=T$ (the \textit{time of maturity} of the option) to buy a share of stock at a fixed \textit{exercise price} $K$. In what follows we restrict to European call options. The question is: how much should one be willing to pay to buy such an option? Let $X_T$ be a \textit{reasonable price}. According to the definition given by Merton,  Black, and Scholes (M-B-S) an investment of this reasonable price in a mixed \textit{portfolio} (i.e part is invested in stock and part in  bond) at time $t=0$, should allow the investor through a \textit{self-financing strategy} (i.e one where the only change in the investor's wealth comes from changes of the prices of the stock and bond) to end up at time $t=T$ with an amount of $(X_T-K)^+:=\max (0,X_T-K)$ which is the same as the payoff, had the option been purchased (cf. \cite{3}). Moreover, such a \textit{reasonable} price allows for no \textit{arbitrage} i.e, it does not allow for risk free unbounded profits. We assume that there are no \textit{transaction costs} and that the portfolio is not made smaller by \textit{consumption}. If $(a_t,b_t), t\in[0,T]$ is a self -financing  \textit{trading strategy} (i.e an amount $a_t$ is invested in stock at time $t$ and an amount $b_t$ is invested in bond at the same time) then the \textit{value} of the portfolio at time $t$ is given by $V_t=a_t\,X_t+b_t\,\beta_t$ where, by the self-financing assumption, $dV_t=a_t\,dX_t+b_t\,d\beta_t$. Here $X_t$ and $\beta_t$ denote, respectively,  the price of the stock and bond at time $t$. We assume that $dX_t= c\,X_t\,dt+\sigma \, X_t \,dB_t$ and $d\beta_t= \beta_t \, r\, dt$ where $B_t$ is classical Brownian motion, $r>0$ is the constant interest rate of the bond, $c>0$ is the \textit{mean rate of return}, and $\sigma >0$ is the \textit{volatility} of the stock. The assets $a_t$ and $b_t$ are in general stochastic processes. Letting $V_t=u(T-t,X_t)$ where $V_T=u(0,X_T)=(X_T-K)^+ $ it can be shown (cf. \cite{3}) that  $u(t,x)$ is the solution of the  Black-Scholes equation

\begin{eqnarray*}
\frac{\partial}{\partial t} u(t,x)  &=&( 0.5\,{\sigma}^2\,x^2\, \frac{{\partial}^2}{\partial x^2} +r\,x\,\frac{\partial}{\partial x} -r)\,u(t,x) \\
u(0,x)&=&(X_T-K)^+ ,\,\,\,\, x>0,\,t\in [0,T]
\end{eqnarray*}

\noindent and it is explicitly given by

\[
u(t,x) =x\,\Phi (g(t,x))-K\,e^{-r\,t}\,\Phi (h(t,x))
\]

\noindent where

\[
g(t,x)= (\ln (x/K) +(r+0.5\,\sigma^2 )\,t ){(\sigma \, \sqrt{t})}^{-1},\,\,\,\,h(t,x)=g(t,x)-\sigma \sqrt{t}
\]

\noindent and

\begin{eqnarray*}
\Phi (x)&=&\frac{1}{\sqrt{2\,\pi}}\,\int_{-\infty}^x\,e^{-y^2/2}\,dy= \frac{1}{2}+\frac{1}{\sqrt{2\,\pi}}\,\sum_{n=0}^{+\infty}\,\frac{(-1)^n}{2^n\,n!}\,\frac{x^{2\,n+1}}{ 2\,n+1 }. \nonumber
\end{eqnarray*}

\noindent Thus a reasonable (in the sense described above) price for a European call option is 

\[
V_0=u(T,X_0) =X_0\,\Phi (g(T,X_0))-K\,e^{-r\,T}\,\Phi (h(T,X_0))
\]

\noindent and the self-financing strategy $(a_t,b_t), t\in[0,T]$ is given by 

\[
a_t=\frac{\partial}{\partial x}u(T-t,X_t),\,\,\,\,
 b_t=\frac{u(T-t,X_t)-a_t\,X_t}{{\beta}_t}.
\]

\section{ Quantum Extension of the M-B-S Model}

 In recent years the fields of Quantum Economics and Quantum Finance have appeared in order to interpret erratic stock  market behavior with the use of quantum mechanical concepts (cf. \cite{Ba1}, \cite{Ba2},\cite{c}-\cite{H}, \cite{M}, and \cite{P1}-\cite{5}). While no approach has yet been proved prevalent, in \cite{5} Segal and Segal introduced quantum effects into the Merton-Black-Scholes model in order to incorporate market features such as the impossibility of simultaneous measurement of prices and their instantaneous derivatives. They did that by adding to the Brownian motion $B_t$ used to represent the evolution of public information affecting the market,  a process $Y_t$ which represents the influence of factors not simultaneously measurable with those involved in $B_t$. They then sketched a calculus for dealing with such processes. Segal and Segal concluded that the combined process $a\,B_t+b\,Y_t$ may be represented as (in their notation) $\Phi\left((a+ib)\,\chi_{[0,t]}\right)$ where for a Hilbert space  element $f$,  $e^{i\,\Phi (f)}$ is  the corresponding  Weyl operator, and $\chi_{[0,t]}$ is the characteristic function of the interval $[0,t]$. In the context of  the  Hudson-Parthasarathy quantum stochastic calculus of \cite{6} and \cite{4} (see Theorem 20.10 of \cite{4}) simple linear combinations of $\Phi (f)$ and  $\Phi (i\,f)$ define the Boson Fock space annihilator and creator operators $A_f$ and $ A_f^{\dagger}$. Segal and Segal used $\Phi (\chi_{[0,t]})$ as the basic integrator process with integrands restricted to a special class of exponential processes. In view of the above reduction of $\Phi$ to $A$ and $ A^{\dagger}$, it makes sense to study option pricing using as integrators the annihilator and creator processes of Hudson-Parthasarathy quantum stochastic calculus, thus exploiting its much larger class of integrable processes than the one considered in \cite{5}.  The Hudson-Parthasarathy calculus has a wide range of applications.  For applications to, for example,  control theory we refer to \cite{01}, \cite{k} and the references therein. Quantum stochastic calculus was designed to describe the dynamics of quantum processes and we propose that we use it to study the non commutative Merton-Black-Scholes model in  the following formulation (notice that our model includes also the Poisson process):  We replace (see \cite{0} for details on quantization) the stock process $\{X_t\,/\,t\geq 0\}$ of the classical Black-Scholes theory by the quantum mechanical process $j_t(X)=U_t^*\,X \otimes 1\,U_t$ where , for each $t\geq0$, $U_t$ is a unitary operator defined on the tensor product  $\mathcal{H} \otimes \Gamma (L^2(\bf{R}_+,\mathcal{ \mathcal{C}  }))$ of a system Hilbert space $\mathcal{H} $ and the noise Boson Fock space $\Gamma=\Gamma  (L^2(\bf{R}_+,\mathcal{ \mathcal{C}  }))$ satisfying 

\begin{equation}
dU_t=-\left(\left(iH+\frac{1}{2}\,L^*L\right)\,dt+ L^* \,S\,dA_t -L\, dA_t^{\dagger}+\left(1-S\right)\,d\Lambda_t\right)U_t,\,\,\,U_0=1\label{e0}
\end{equation}

\noindent where $X>0$,  $H$, $L$, $S$ are in $\mathcal{B}(\mathcal{H})$, the space of bounded linear operators on $\mathcal{H} $, with $S$ unitary and $X,\,H$ self-adjoint.   We identify time-independent, bounded, system space operators $x$ with their ampliation $x \otimes 1$ to $\mathcal{H} \otimes \Gamma(L^2(\bf{R}_+, \mathcal{\mathcal{C}}))$. The value process $V_t$ is defined for $t\in[0,T]$ by $V_t= a_t\, j_t(X)  +b_t\,\beta_t$ with terminal condition $V_T= ( j_T(X) -K)^+=\max (0,  j_T(X)   -K)$, where $K>0$ is a bounded self-adjoint system operator corresponding to the strike price of the quantum option,  $a_t$ is a real-valued function,  $b_t$ is in  general an observable quantum stochastic processes (i.e $b_t$ is a self-adjoint operator for each $t\geq 0$)  and $\beta_t= \beta_0\,e^{  t\,r }$ where $\beta_0$ and $r$ are positive  real numbers. Therefore $ b_t =(V_t- a_t\, j_t(X))\,\beta_t^{-1}$. We interpret the above in the sense of expectation i.e given $u\otimes \psi (f)$  in the exponential domain of $\mathcal{H} \otimes \Gamma$, where we will always assume $u \neq 0$ so that $\|u\otimes \psi (f) \| \neq 0$,

\begin{eqnarray*}
<u\otimes \psi (f),V_t\,u \otimes \psi (f)>&=& a_t\,<u\otimes \psi (f), j_t(X)\,u \otimes \psi (f)> \\
& +&<u\otimes \psi (f),b_t\,u \otimes \psi (f)> \,\beta_t
\end{eqnarray*}

\noindent (i.e the value process is always in reference to a particular quantum mechanical state, so we can eventually reduce to real numbers)  and 

\begin{eqnarray*}
<u\otimes \psi (f),V_T\,u \otimes \psi (f)> &=& <u\otimes \psi (f),( j_T(X) -K)^+\,u \otimes \psi (f)>\\
&=&\max (0,<u\otimes \psi (f), ( j_T(X)   -K)\,u \otimes \psi (f)>).
\end{eqnarray*}
 
\noindent As in the classical case we assume that the portfolio $(a_t,b_t), t\in[0,T]$ is  self -financing  i.e 

\[
dV_t=a_t\,dj_t(X)+b_t\,d\beta_t
\]

\noindent or equivalently

\[
da_t\,\cdot\,j_t(X)+da_t\,\cdot\,dj_t(X)+db_t\,\cdot\,\beta_t+db_t\,\cdot\,d\beta_t=0.
\]

\begin{remark}\end{remark} The fact that the value process (and all other operator processes $X_t$ appearing in this paper) is always in reference to a particular quantum mechanical state, allows for a direct translation of all classical financial concepts described in Section 1 to the quantum case by considering the expectation (or matrix element) $<u\otimes \psi (f),X_t\,u \otimes \psi (f)>$ of the process at each time $t$. If the process is classical (i.e, if $X_t\in\mathbb{R}$) then we may divide out 
$\|u \otimes \psi (f)\|^2$ and everything is reduced to the classical case described in Section 1.  

 \begin{lemma}  Let $j_t(X)=U_t^*\,X \otimes 1\,U_t$ where $\{U_t\,/\,t\geq 0\}$ is the solution of (\ref{e0}).  If

 \[
\alpha = [L^*,X]\,S,\,\,\, {\alpha}^{\dagger}=S^*\,[X,L],\,\,\,  \lambda = S^*\,X\,S-X,
\]

\noindent and 

\[
\theta=i\,[H,X]-\frac{1}{2}\,\{   L^*\,L\,X+X\,L^*\,L-2\,L^*\,X\,L\}
\]

\noindent then

\begin{eqnarray}
&dj_t(X)=j_t({\alpha}^{\dagger})\,dA^{\dagger}_t+j_t(\lambda )\,d\Lambda_t  +   j_t(\alpha)\,dA_t   +j_t( \theta )\,dt &\label{e1}
\end{eqnarray}

\noindent and for $k \geq 2$

\begin{eqnarray}
&\left( dj_t(X) \right) ^k=j_t({\lambda}^{k-1}\, {\alpha}^{\dagger})\,dA^{\dagger}_t+j_t({\lambda}^{k} )\,d\Lambda_t  +   j_t(\alpha {\lambda}^{k-1})\,dA_t   +j_t( \alpha   \, {\lambda}^{k-2}   \,  {\alpha}^{\dagger})\,dt &\label{e2}
\end{eqnarray}

\end{lemma}

\begin{proof} Equation (\ref{e1}) is a standard result of quantum flows theory (cf. \cite{ 4}). To prove  (\ref{e2}) we notice that for $k=2$, using (\ref{e1}), the It\^{o} table

\begin{center}
\begin{tabular}{c|cccc}
$\cdot$&$dA_t^{\dagger}$&$d\Lambda_t$&$dA_t$&$dt$\\
\hline
$dA_t^{\dagger}$&$0$&$0$&$0$&$0$\\
$d\Lambda_t$&$dA_t^{\dagger}$&$d\Lambda_t$&$0$&$0$\\
$dA_t$&$dt$&$dA_t$&$0$&$0$\\
$dt$&$0$&$0$&$0$&$0$
\end{tabular}
\end{center}

\noindent and the homomorhism property $j_t(x\,y)=j_t(x)\, j_t(y) $, we obtain

\begin{eqnarray*}
(dj_t(X) )^2 = dj_t(X)\,d j_t(X)=  j_t(\lambda \, {\alpha}^{\dagger})\, dA^{\dagger}_t+j_t({\lambda}^{2} )\, d\Lambda_t  +   j_t( \alpha \lambda ) \, dA_t   + j_t ( \alpha  \,  { \alpha }^{\dagger}) \, dt \nonumber
\end{eqnarray*}

\noindent so (\ref{e2})  is true for $k=2$. Assuming (\ref{e2}) to be true for $k$ we have

\begin{eqnarray*}
&&(dj_t(X) )^{k+1} =  dj_t(X) \, (dj_t(X))^k \nonumber \\
\noalign{\vskip .05 true in}
&& \quad =
dj_t(X) \,\left(  j_t({\lambda}^{k-1}\, {\alpha}^{\dagger})\,dA^{\dagger}_t+j_t({\lambda}^{k} )\,d\Lambda_t  +   j_t(\alpha {\lambda}^{k-1})\,dA_t   +j_t( \alpha   \, {\lambda}^{k-2}   \,  {\alpha}^{\dagger})\,dt \right)\nonumber \\
\noalign{\vskip .05 true in}
&& \quad =
 j_t({\lambda}^{k}\, {\alpha}^{\dagger})\,dA^{\dagger}_t+j_t({\lambda}^{k+1} )\,d\Lambda_t  +   j_t(\alpha {\lambda}^{k})\,dA_t   +j_t( \alpha   \, {\lambda}^{k-1}   \,  {\alpha}^{\dagger})\,dt \nonumber
\end{eqnarray*}

\noindent Thus (\ref{e2}) is true for $k+1$ also.

\end{proof}

\section{ Derivation of the Quantum Black-Scholes Equation}

In the spirit of the previous section, let $ V_t:=F(t,j_t(X)) $ where $F:[0,T] \times \mathcal{B}(\mathcal{H} \otimes \Gamma ) \longrightarrow \mathcal{B} ( \mathcal{H} \otimes \Gamma)$  is the extension to self-adjoint operators $x=j_t(X) $  of the analytic function  $F(t,x)= \sum_{n,k=0}^{+\infty}\,a_{n,k}(t_0,x_0)\,(t-t_0)^n\,(x-x_0)^k$, where $x$ and $a_{n,k}(t_0,x_0)$  are in $ \bf{C}$, and for $\lambda,\mu \in \{0,1,...\}$ 

\begin{eqnarray*}
&&F_{\lambda\,\mu}(t,x):=\frac{\partial^{\lambda + \mu} F}{\partial t^{\lambda}\,\partial x^{\mu}}(t,x)\\
\noalign{\vskip .05 true in}
&& \quad =
\sum_{n=\lambda,k=\mu}^{+\infty}\,\frac{ n! }{ (n-\lambda)! }\,\frac{ k!  }{ (k -\mu )!  }\,a_{n,k}(t_0,x_0)\,(t-t_0)^{n-\lambda}\,(x-x_0)^{k-\mu}\nonumber
\end{eqnarray*}

\noindent and so, if $1$ denotes the identity operator then 

\[
a_{n,k}(t_0,x_0)=a_{n,k}(t_0,x_0)\,1=\frac{ 1 }{ n!\,k! }\,F_{n\,k}(t_0,x_0).
\]

\noindent Notice that for $(t_0,x_0)=(0,0)$ we have

\[
V_t= \sum_{n,k=0}^{+\infty}\,a_{n,k}(0,0)\,t^n\,j_t(X)^k 
= \sum_{n,k=0}^{+\infty}\,a_{n,k}(0,0)\,t^n\,j_t(X^k).
\]

\begin{proposition} (Quantum Black-Scholes Equation)  

\begin{eqnarray*}
 &a_{1,0}(t,j_t(X))+a_{0,1}(t,j_t(X))\,j_t( \theta )
+\sum_{k=2}^{+\infty}\,a_{0,k}(t,j_t(X))\,j_t(\alpha \, {\lambda}^{k-2}   \,  {\alpha}^{\dagger}) =&\\
&  a_t \, j_t( \theta )+V_t\,r-a_t\, j_t(X)\,r&
\end{eqnarray*}

\noindent (this is the quantum analogue of the classical Black-Scholes equation) and

\begin{eqnarray*}
a_{0,1}(t,j_t(X))\,j_t({\alpha}^{\dagger})+\sum_{k=2}^{+\infty}\,a_{0,k}(t,j_t(X))\,j_t({\lambda}^{k-1}\, {\alpha}^{\dagger}) &=&  a_t\,j_t({ \alpha }^{ \dagger } )\\
 a_{0,1}(t,j_t(X))\,j_t(\alpha )+\sum_{k=2}^{+\infty}\,a_{0,k}(t,j_t(X))\,j_t({\alpha \,\lambda}^{k-1})&=& a_t\,j_t( \alpha )\\
\sum_{k=1}^{+\infty}\,a_{0,k}(t,j_t(X))\,j_t({\lambda}^{k} ) &=&  a_t\, j_t( \lambda ).
\end{eqnarray*}

\end{proposition}

\begin{proof} By Lemma 2.1 and the It\^{o} table for quantum stochastic differentials

\begin{eqnarray*}
dV_t&=&dF(t,j_t(X) )= F(t+dt,j_{t+dt}(X) )- F(t,j_t(X) )\\
&=&F(t+dt,j_t(X)+dj_t(X) )- F(t,j_t(X) )\nonumber \\
&=&\sum_{\stackrel{n,k=0}{n+k>0}}^{+\infty}\,a_{n,k}(t,j_t(X))\,(dt)^n\,( dj_t(X))^k\\
&=&a_{1,0}(t,j_t(X))\,dt+\sum_{k=1}^{+\infty}\,a_{0,k}(t,j_t(X))\,( dj_t(X))^k\nonumber\\
&=&a_{1,0}(t,j_t(X))\,dt+a_{0,1}(t,j_t(X))\,dj_t(X)+\sum_{k=2}^{+\infty}\,a_{0,k}(t,j_t(X))\,\{j_t({\lambda}^{k-1}\, {\alpha}^{\dagger})\,dA^{\dagger}_t\\
&+&j_t({\lambda}^{k} )\,d\Lambda_t  +   j_t(\alpha {\lambda}^{k-1})\,dA_t+j_t( \alpha   \, {\lambda}^{k-2}   \,  {\alpha}^{\dagger})\,dt  \}\nonumber
\end{eqnarray*}

\noindent where $\alpha , {\alpha}^{\dagger}, \lambda$ are as in Lemma 2.1. Thus

\begin{eqnarray}
dV_t&=&\left(a_{1,0}(t,j_t(X))+a_{0,1}(t,j_t(X))\,j_t( \theta )+\sum_{k=2}^{+\infty}\,a_{0,k}(t,j_t(X))\,j_t(\alpha   \, {\lambda}^{k-2}   \,  {\alpha}^{\dagger})\right) \,dt\nonumber\\
&+&\left( a_{0,1}(t,j_t(X))\,j_t({\alpha}^{\dagger})+\sum_{k=2}^{+\infty}\,a_{0,k}(t,j_t(X))\,j_t({\lambda}^{k-1}\, {\alpha}^{\dagger})\right)\,dA^{\dagger}_t\nonumber\\
&+&\left( a_{0,1}(t,j_t(X))\,j_t(\alpha )+\sum_{k=2}^{+\infty}\,a_{0,k}(t,j_t(X))\,j_t({\alpha \,\lambda}^{k-1})\right)\,dA_t\nonumber\\
&+&\sum_{k=1}^{+\infty}\,a_{0,k}(t,j_t(X))\,j_t({\lambda}^{k} )\,d\Lambda_t \label{e9}
\end{eqnarray}

\noindent where $\theta$ is as in Lemma 2.1. We can obtain another expression for $dV_t$  with the use of the self-financing property. We have

\begin{eqnarray*}
dV_t&=&a_t\,dj_t(X)+b_t\,d\beta_t=a_t\,dj_t(X)+b_t\,\beta_t\,r\,dt\\
&=&a_t\,dj_t(X)+(V_t- a_t\, j_t(X))\,\beta_t^{-1}\,\beta_t\,r\,dt\nonumber\\
&=&a_t\,dj_t(X)+(V_t- a_t\, j_t(X))\,r\,dt\nonumber\\
&=&a_t\,\left( j_t({\alpha}^{\dagger})\,dA^{\dagger}_t+j_t(\lambda )\,d\Lambda_t  +   j_t(\alpha)\,dA_t   +j_t( \theta )\,dt \right)+(V_t- a_t\, j_t(X))\,r\,dt\nonumber
\end{eqnarray*}

\noindent which can be written as

\begin{eqnarray}
dV_t &=& (  a_t \, j_t( \theta )+
V_t\,r-a_t\, j_t(X)\,r ) \,dt+ a_t\,j_t({ \alpha }^{ \dagger } )\, dA^{\dagger }_t+a_t\,j_t( \alpha )\,dA_t\nonumber\\
&+&a_t\, j_t( \lambda )\, d\Lambda_t \label{e11}
\end{eqnarray}

\noindent Equating the coefficients of $dt$ and the quantum stochastic differentials in 
(\ref{e9}) and (\ref{e11}) we obtain the desired equations.

\end{proof}

\section{The case $S=1$: Quantum Brownian motion}

 \begin{proposition}  Let $F$ be as in the previous section. If   $S=1$ then the equations  of  Proposition 3.1 combine into  

\[
u_{1\,0}(t,x)=\frac{1}{2}\,u_{0\,2}(t,x)\,g(x)+ u_{0\,1}(t,x)\,h(x) -u(t,x)\,r\]

\noindent with initial condition $ u (0,j_T(X))=( j_T(X) -K)^+$ where $ u(t,x)=F(T-t,x)$, $g(x) = [ y^*,x]\, [x,y]$, $h(x)=x\,r$ and $x,y\in \mathcal{B}(\mathcal{H} \otimes \Gamma )  $

\end{proposition}

\begin{proof} If $S=1$ then, in the notation of Lemma 2.1, $\alpha = [L^*,X]$,   ${\alpha}^{\dagger}=[X,L]$, $\lambda = 0$, and $\theta=i\,[H,X]-\frac{1}{2}\,\{   L^*\,L\,X+X\,L^*\,L-2\,L^*\,X\,L\}$ and  the equations of Proposition  3.1 reduce to 

\[
a_{1,0}(t,j_t(X))+a_{0,1}(t,j_t(X))\,j_t( \theta )+a_{0,2}(t,j_t(X))\,j_t(\alpha   \,  {\alpha}^{\dagger}) =  a_t \, j_t( \theta )+V_t\,r-a_t\, j_t(X)\,r
\]

\noindent and

\begin{eqnarray*}
a_{0,1}(t,j_t(X))\,j_t({\alpha}^{\dagger}) &=&  a_t\,j_t({ \alpha }^{ \dagger } )\\
 a_{0,1}(t,j_t(X))\,j_t(\alpha )&=& a_t\,j_t( \alpha ) 
\end{eqnarray*}

 \noindent which are condensed into

\[
 a_{1,0}(t,j_t(X))+a_{0,1}(t,j_t(X))\,j_t( \theta )+a_{0,2}(t,j_t(X))\,j_t(\alpha   \,  {\alpha}^{\dagger}) =  a_t \, j_t( \theta ) +V_t\,r-a_t\, j_t(X)\,r
\]

\noindent and

\[ 
a_{0,1}(t,j_t(X))=  a_t.
\]

\noindent Upon substituting the second of the last two equations into the first one and simplifying we obtain 

\[
a_{1,0}(t,j_t(X))+a_{0,2}(t,j_t(X))\,j_t(   [L^*,X]   \,  [X,L]) +  a_{0,1}(t,j_t(X)) \, j_t( X )\,r-V_t\,r=0 
\]

\noindent which can be written as 

\[
 F_{1\,0}(t,j_t(X))+\frac{1}{2}\,F_{0\,2}(t,j_t(X))\,j_t([L^*,X]\,[X,L]) + F_{0\,1}(t,j_t(X)) \, j_t( X )\,r=F (t, j_t(X))\,r 
\]

\noindent with terminal condition $F (T,j_T(X))=( j_T(X) -K)^+$. Letting  $x = j_t(X)$, $y= j_t(L)$ be arbitrary elements in $\mathcal{B}(\mathcal{H} \otimes \Gamma )  $ and letting  $g(x) = [ y^*,x]   \,  [x,y]$, $h(x)=x\,r$,  we obtain

\[
F_{1\,0}(t,x)+\frac{1}{2}\,F_{0\,2}(t,x)\,g(x)+ F_{0\,1}(t,x)\,h(x) =F(t,x)\,r.
\]

 \noindent  Letting  $u(t,x):=F(T-t,x)$,  $u_{1\,0}(t,x)=-F_{1\,0}(T-t,x)$, $u_{0\,2}(t,x)=F_{0\,2}(T-t,x)$ and $u_{0\,1}(t,x)=F_{0\,1}(T-t,x)$  we obtain

\[
 u_{1\,0}(t,x)=\frac{1}{2}\,u_{0\,2}(t,x)\,g(x)+ u_{0\,1}(t,x)\,h(x) -u(t,x)\,r
\]

\noindent with $u (0,j_T(X))=( j_T(X) -K)^+$. 

\end{proof}

\section{ The case $S\neq1$: Quantum Poisson Process}

 In this section we examine the equations of Proposition  3.1  under the assumption $S\neq1$.

\begin{proposition} Let $F$ be as in Section 3. If    $[X,S] = S $ then the equations of Proposition  3.1 combine into 

\[
u_{1\,0}(t,x)=\sum_{k=2}^{+\infty}\,\frac{1}{k!}\,u_{0\,k}(t,x)\,g(x)+ u_{0\,1}(t,x) \,h(x)-u (t,x)\,r
\]

 \noindent with initial condition $ u (0,j_T(X))=( j_T(X) -K)^+$ where $u(t,x)=F(t-T,x)$, $g(x) = [ y^*,x]   \,  [x,y] -i\,[z,x]+\frac{1}{2}\,\{y^*\,y\,x+x\,y^*\,y-2\,y^*\,x\,y\}$, $h(x)=x\,r$ and $x,y,z\in \mathcal{B}(\mathcal{H} \otimes \Gamma )  $

\end{proposition}

\begin{proof}  Since $X$ is self-adjoint and $S$ is unitary,  assuming that $[X,S] = S$ is equivalent to assuming that $\lambda = S^*\,X\,S-X=1 $ and  the equations of Proposition  3.1 take the form

\[
 a_{1,0}(t,j_t(X))+a_{0,1}(t,j_t(X))\,j_t( \theta )
+\sum_{k=2}^{+\infty}\,a_{0,k}(t,j_t(X))\,\,j_t(\alpha\,{\alpha}^{\dagger}) =  a_t \, j_t( \theta )+V_t\,r-a_t\, j_t(X)\,r
\]

\noindent and

\begin{eqnarray*}
a_{0,1}(t,j_t(X))\,j_t({\alpha}^{\dagger})+\sum_{k=2}^{+\infty}\,a_{0,k}(t,j_t(X))\,\,j_t( {\alpha}^{\dagger}) &=&  a_t\,j_t({ \alpha }^{ \dagger } )\label{e132} \\
 a_{0,1}(t,j_t(X))\,j_t(\alpha )+\sum_{k=2}^{+\infty}\,a_{0,k}(t,j_t(X))\,\,j_t(\alpha )&=& a_t\,j_t( \alpha )\label{e142} \\
\sum_{k=1}^{+\infty}\,a_{0,k}(t,j_t(X)) &=&  a_t\label{e152}
\end{eqnarray*}

\noindent which are satisfied if

\[
 a_{1,0}(t,j_t(X))+a_{0,1}(t,j_t(X))\,j_t( \theta )
+\sum_{k=2}^{+\infty}\,a_{0,k}(t,j_t(X))\,\,j_t(\alpha\,{\alpha}^{\dagger}) = a_t \, j_t( \theta )+V_t\,r-a_t\, j_t(X)\,r
\]

\noindent and $a_t=\sum_{k=1}^{+\infty}\,a_{0,k}(t,j_t(X)) $ which,  if substituted in the previous one,  yields

\[
 a_{1,0}(t,j_t(X))+a_{0,1}(t,j_t(X))\,j_t( X )\,r
+\sum_{k=2}^{+\infty}\,a_{0,k}(t,j_t(X))\,\,\left(j_t(\alpha\,{\alpha}^{\dagger}-\theta )+ j_t(X)\,r\right)=V_t\,r.
\]

\noindent But

\begin{eqnarray*}
 j_t(\alpha\,{\alpha}^{\dagger}-\theta )&=& [j_t(L)^*,j_t(X)]\, [j_t(X),j_t(L)]-i\,[j_t(H),j_t(X)]   \label{e125} \\ 
&+&\frac{1}{2}\,\{  j_t( L)^*\,j_t(L)\,j_t(X)+j_t(X)\,j_t(L)^*\,j_t(L)-2\,j_t(L)^*\,j_t(X)\,j_t(L)\} \nonumber
\end{eqnarray*}

\noindent Letting $x = j_t(X)$, $y= j_t(L)$, $z=j_t(H)$, $h(x)=x\,r$  and

\begin{eqnarray*}
g(x) &=& [ y^*,x]   \,  [x,y] -i\,[z,x]+\frac{1}{2}\,\{y^*\,y\,x+x\,y^*\,y-2\,y^*\,x\,y\}
\end{eqnarray*}

\noindent using the notation of the previous section we obtain the Black-Scholes equation for the  case $S\neq 1$ as stated in the Proposition.

\end{proof}

\section{ Solution of the Quantum Brownian Motion Black-Scholes Equation}

 To solve the Quantum Brownian motion Black-Scholes equation  we assume that $  j_t( X^2 )= j_t([L^*,X]\, [X,L])$ which is the same as $ X^2=  [L^*,X]   \,  [X,L]  $. 
Since $X=X^*$, it follows that  $[L^*,X]= [X,L]^*$  and so letting $\phi (X)=[X,L]$   we find $  X^2= \phi (X)^* \,  \phi (X) $ i.e  $\phi (X) =W\,X $ which implies that $[X,L] = W\,X  $ and $[L^*, X ]=  X\,W^*$, where $W$ is an arbitrary  unitary operator acting on the system space. In this case equation (\ref{e1}) takes the form

 \[
dj_t(X)=j_t\left( i\,[H,X]+\frac{1}{2}\,\left( L^*\,W\,X+X\,W^*\,L   \right)  \right)\,dt   +j_t(   X  \,W)\,dA^{\dagger}_t+ j_t(W^* \, X )\,dA_t.
\]

 \begin{lemma}  If $H>0$ is a bounded self-adjoint operator on a Hilbert space $\mathcal{H}$ then there exists  a bounded self-adjoint operator $A$ on  $\mathcal{H}$ such that $H=e^A$.
\end{lemma}

\begin{proof} Let $H=\int_a^b \,\lambda\,dE_{\lambda}$ where $[a,b]\subset (0,+\infty)$ and $a\leq \|H\|\leq b$. Letting  $\lambda =e^{\mu}$ we obtain $H=\int_{\ln a}^{\ln b} \,e^{\mu}\,dF(\mu)$ where $F(\mu)=E(e^{\mu})$. Thus $H=e^A$ where $A=\int_{\ln a}^{\ln b} \,\mu\,dF(\mu)$ with $ \|A\|\leq \max \,(|\ln a|,|\ln b|)$. To show that the family $\{F(\mu)/  \ln a\leq \mu \leq \ln b\}$ is a resolution of the identity we notice that for $h\in\mathcal{H}$ and $\lambda, \mu \in [\ln a,\ln b]$ we have:

\begin{eqnarray*}
 &(i)&  F(\lambda)\,F(\mu )=E(e^{\lambda})\,E(e^{\mu})=E(e^{\lambda} \wedge  e^{\mu})=F(\lambda\wedge \mu), \\
&(ii)& \lim_{\lambda \rightarrow \mu^-}F(\lambda)\,h=\lim_{e^{\lambda} \rightarrow e^{{\mu}^-}}E(e^{\lambda})\,h=E(e^{\mu})\,h=F(\mu)\,h,\\
 &(iii)&  \lambda<\mu\Rightarrow e^{\lambda} <e^{\mu}\Rightarrow E(e^{\lambda})< E(e^{\mu})\Rightarrow F(\lambda)<F(\mu),\\
 &(iv)&   \lambda < \ln a \Rightarrow e^{\lambda} < a\Rightarrow 
  E(e^{\lambda})=0    \Rightarrow F(\lambda)=0,\\ 
&(v) &  \lambda > \ln b \Rightarrow e^{\lambda} > b \Rightarrow E(e^{\lambda })=1    \Rightarrow  F(\lambda )=1.
\end{eqnarray*}

and the proof is complete.

\end{proof}

\bigskip

\noindent The equation in Proposition 4.1 now has the form 

\[
u_{1\,0}(t,x)=\frac{1}{2}\,u_{0\,2}(t,x)\,x^2+ u_{0\,1}(t,x)\,x\,r -u(t,x)\,r
\]

\noindent with initial condition  $ u (0,j_T(X))=( j_T(X) -K)^+$ where we may assume that $x$ is a bounded self-adjoint operator. Since 

\[
u(t,x)=F(T-t,x)= \sum_{n,k=0}^{+\infty}\,a_{n,k}(0,0)\,(T-t)^n\,x^k
\]

\noindent and   $x=j_t(X)>0$, and $K$ are invertible,  we may let  $x=K\,e^z$ where $z$ is a bounded self-adjoint operator commuting with $K$, and obtain

\begin{eqnarray*}
\omega (t,z)&:=&u(t,K\,e^z)= \sum_{n,k=0}^{+\infty}\,a_{n,k}(0,0)\,(T-t)^n\,(K\,e^z)^k\label{C}\\
\omega_{0\,1} (t,z)&=& \sum_{n,k=0}^{+\infty}\,a_{n,k}(0,0)\,(T-t)^n\,k\,(K\,e^z)^k=\sum_{n=0,k=1}^{+\infty}\,a_{n,k}(0,0)\,(T-t)^n\,k\,x^{k}\\
 &=&\sum_{n=0,k=1}^{+\infty}\,a_{n,k}(0,0)\,(T-t)^n\,k\,x^{k-1}\,\,x = u_{0\,1} (t,x) \,\,x \nonumber
\end{eqnarray*}

\noindent Similarly

\begin{eqnarray*}
\omega_{0\,2} (t,z)&=& \sum_{n=0,k=1}^{+\infty}\,a_{n,k}(0,0)\,(T-t)^n\,k^2\,(K\,e^z)^k=\sum_{n=0,k=1}^{+\infty}\,a_{n,k}(0,0)\,(T-t)^n\,k^2\,x^{k} \nonumber\\
&=&\sum_{n=0,k=1}^{+\infty}\,a_{n,k}(0,0)\,(T-t)^n\,\left(k\,(k-1)+k\right)\,x^{k}\\
& =&\sum_{n=0,k=2}^{+\infty}\,a_{n,k}(0,0)\,(T-t)^n\,k\,(k-1)\,\,x^{k-2}\,\,x^2\\
&+& \sum_{n=0,k=1}^{+\infty}\,a_{n,k}(0,0)\,(T-t)^n\,k\,\,x^{k-1}\,\,x\\
&=&u_{0\,2} (t,x) \,\,x^2+ u_{0\,1} (t,x) \,\,x \nonumber
\end{eqnarray*}

\noindent and so 

\[
\omega_{0\,2} (t,z)-\omega_{0\,1} (t,z)=u_{0\,2} (t,x) \,\,x^2.\]

\noindent  Finally

\begin{eqnarray*}
\omega_{1\,0} (t,z)&=& -\sum_{n=1,k=0}^{+\infty}\,a_{n,k}(0,0)\,n\,(T-t)^{n-1}\,(K\,e^z)^k\\
&=&-\sum_{n=1,k=0}^{+\infty}\,a_{n,k}(0,0)\,n\,(T-t)^{n-1}\,x^{k} =u_{1\,0} (t,x) \nonumber
\end{eqnarray*}

\noindent and so

\begin{eqnarray}
 &\omega_{1\,0}(t,z)=\frac{1}{2}\,\omega_{0\,2}(t,z)+ \omega_{0\,1}(t,z)\,\left(r-\frac{1}{2}\right) -\omega(t,z)\,r& \label{e20099}
\end{eqnarray}

\noindent with initial condition $\omega(0,z_T)=( j_T(X) -K)^+$ where   $z_T$ is defined by $K\,e^{z_T}= j_T(X)$.

 \begin{theorem} In analogy with  the classical case presented in section 1, the solution of  (\ref{e20099}) is given by 

\begin{eqnarray*}
\omega (t,z) =K\,e^z\,\,\Phi (g(t,K\,e^z))-K\,\Phi (h(t,K\,e^z))\,e^{-r\,t}
\end{eqnarray*}

\noindent  where 

\begin{eqnarray*}
g(t,  K\,e^z )&=&  z\,t^{-1/2}+(r+0.5)\,t^{1/2}\\
h(t,K\,e^z)&=& z\,t^{-1/2}+(r-0.5)\,t^{1/2}, 
\end{eqnarray*}

\noindent  and 

\begin{eqnarray*}
\Phi (x)&=& \frac{1}{2}+\frac{1}{\sqrt{2\,\pi}}\,\sum_{n=0}^{+\infty}\,\frac{(-1)^n}{2^n\,n!}\,\frac{x^{2\,n+1}}{ 2\,n+1 } 
\end{eqnarray*}

\end{theorem}

\begin{proof} We have

\[
\omega _{1\,0}(t,z) =K\,e^z\,\,(\Phi \circ g)_{1\,0}(t,K\,e^z)-K\, (\Phi \circ h)_{1\,0}  (t,K\,e^z)\,e^{-r\,t}+K\,(\Phi \circ h) (t,K\,e^z)\,r\,e^{-r\,t},
\]

\[
\omega _{0\,1}(t,z) =K\,e^z\,\,(\Phi \circ g)(t,K\,e^z)+K\,e^z\,\,(\Phi \circ g)_{0\,1}(t,K\,e^z)-K\, (\Phi \circ h)_{0\,1}  (t,K\,e^z)\,e^{-r\,t},
\]

\noindent and

\begin{eqnarray*}
\omega _{0\,2}(t,z)& =&K\,e^z\,\,(\Phi \circ g)(t,K\,e^z)+2\,K\,e^z\,\,(\Phi \circ g)_{0\,1}(t,K\,e^z)+ K\, (\Phi \circ g)_{0\,2}  (t,K\,e^z)\nonumber\\
&-&K\, (\Phi \circ h)_{0\,2}  (t,K\,e^z)\,e^{-r\,t}\nonumber
\end{eqnarray*}

\noindent where

\begin{eqnarray*}
(\Phi \circ h) (t,K\,e^z)&=& \frac{1}{2}+\frac{1}{\sqrt{2\,\pi}}\,\sum_{n=0}^{+\infty}\,\frac{(-1)^n}{2^n\,n!}\,\frac{(  z\,t^{-1/2}+(r-0.5)\,t^{1/2} )^{2\,n+1}}{ 2\,n+1 }\label{c1} \\
(\Phi \circ g) (t,K\,e^z)&=& \frac{1}{2}+\frac{1}{\sqrt{2\,\pi}}\,\sum_{n=0}^{+\infty}\,\frac{(-1)^n}{2^n\,n!}\,\frac{(  z\,t^{-1/2}+(r+0.5)\,t^{1/2} )^{2\,n+1}}{ 2\,n+1 }\label{d1} \\
\end{eqnarray*}

\noindent Thus

\begin{eqnarray*}
\omega _{1\,0}(t,z)-\frac{1}{2}\,\omega _{0\,2}(t,z) - \omega _{0\,1}(t,z)\,(r-\frac{1}{2})+\omega (t,z)\,r& =&K\,\left(A\,e^{-r\,t}+e^z\,B\right)
\end{eqnarray*}

\noindent where

\begin{eqnarray*}
A&=&-(\Phi \circ h)_{1\,0}(t,K\,e^z)    +\frac{1}{2}\,  (\Phi \circ h)_{0\,2}(t,K\,e^z) +(\Phi \circ h)_{0\,1}(t,K\,e^z)\,(r  -\frac{1}{2})\nonumber \\
&&\\
B&=&(\Phi \circ g)_{1\,0} (t,K\,e^z)    -\frac{1}{2}\, (\Phi \circ g)_{0\,2}  (t,K\,e^z)-(\Phi \circ g)_{0\,1}(t,K\,e^z)\,(r+\frac{1}{2}) \nonumber
\end{eqnarray*}

\noindent It follows that $A=B=0$ thus proving (\ref{e20099}). Moreover, in order to prove that the initial condition is satisfied, we have

\begin{eqnarray*}
 \omega (0,z_T) &=&K\,e^{z_T}\,\Phi (g(0,K\,e^{z_T}))-K\,\Phi (h(0,K\,e^{z_T}))\\
&=&\left(K\,e^{z_T}-K\right)\,\Phi (g(0,K\,e^{z_T}))+K\,\left(\Phi (g(0,K\,e^{z_T}))-  \Phi (h(0,K\,e^{z_T})) \right).
\end{eqnarray*}

 \noindent But 

\begin{eqnarray*}
g(0,K\,e^{z_T})-  h(0,K\,e^{z_T}) =\lim_{t\rightarrow 0^+}\left( \frac{z}{\sqrt{t}}+(r+0.5)\,\sqrt{t}-\frac{z}{\sqrt{t}}-(r-0.5)\,\sqrt{t}\right)=0
\end{eqnarray*}

\noindent and so $\Phi (g(0,K\,e^{z_T}))-  \Phi (h(0,K\,e^{z_T})) =0$. Thus,  it suffices to show that

\begin{eqnarray*}
\Phi (g(0,K\,e^{z_T}))=\left\{
\begin{array}{cc}1 &\mbox{ if } K\,e^{z_T}\geq K \\
 0 &\mbox{ if } K\,e^{z_T} < K 
\end{array}\right.
\end{eqnarray*}

\noindent We have

\begin{eqnarray*}
\Phi ( g (0,K\,e^{z_T}))&=&\lim_{t\rightarrow 0^+}\,(\Phi \circ g) (t,K\,e^{z_T}))\\
&=&\frac{1}{2}+\lim_{t\rightarrow 0^+}\,\frac{1}{\sqrt{2\,\pi}}\,\sum_{n=0}^{+\infty}\,\frac{(-1)^n}{2^n\,n!}\,  \frac{1}{ t^{n+1/2}}\,\frac{  z_T^{2\,n+1}}{ 2\,n+1 }\nonumber
\end{eqnarray*}

\noindent Suppose that  $K\,e^{z_T}\geq K$. Then $z_T\geq 0$ and by the spectral resolution theorem
 $z_T^{2\,n+1}=\int_0^{+\infty} \,{\lambda}^{2\,n+1}\,dE_{\lambda}$. So

\begin{eqnarray*}
\Phi ( g (0,K\,e^{z_T}))&=&\frac{1}{2}+\lim_{t\rightarrow 0^+}\,\frac{1}{\sqrt{2\,\pi}}\,\sum_{n=0}^{+\infty}\,\frac{(-1)^n}{2^n\,n!}\,  \frac{1}{ t^{n+1/2}}\,\int_0^{+\infty}\,\frac{  \lambda^{2\,n+1}}{ 2\,n+1 }\,dE_{\lambda}\\
&=&\frac{1}{2}+\lim_{t\rightarrow 0^+}\,\frac{1}{\sqrt{2\,\pi}}\,\int_0^{+\infty}\,\int_0^{\frac{\lambda}{ \sqrt{t}}  }\,e^{-\frac{s^2}{2}}\,ds\,\,dE_{\lambda}\nonumber\\
&=&\frac{1}{2}+\frac{1}{\sqrt{2\,\pi}}\,\int_0^{+\infty}\,\int_0^{ +\infty }\,e^{-\frac{s^2}{2}}\,ds\,\,dE_{\lambda}\nonumber\\
&=&\frac{1}{2}+\frac{1}{\sqrt{2\,\pi}}\,\int_0^{+\infty}\,\frac{ \sqrt{2\,\pi} }{2}\,\,dE_{\lambda}=1\nonumber
\end{eqnarray*}

\noindent Similarly, if  $K\,e^{z_T} < K$ then $z_T < 0$ and if we let $z_T=-w_T$ where $w_T=\int_0^{+\infty} \,\lambda\,dE_{\lambda}\, > 0$, then 

\[
z_T^{2\,n+1}=(-1)^{2\,n+1}\,\int_0^{+\infty} \,{\lambda}^{2\,n+1}\,dE_{\lambda}=-\int_0^{+\infty} \,{\lambda}^{2\,n+1}\,dE_{\lambda}
\]

\noindent  and so, as before, $\Phi ( g (0,K\,e^{z_T}))=\frac{1}{2}-\frac{1}{2}\,\cdot\,1=0$.

\end{proof}

 \begin{corollary} The \textit{reasonable} price for a quantum option is $\omega (T,z_0)$ where $\omega$ is as in Theorem 6.1 and $z_0$ is defined by $X=K\,e^{z_0}$. The associated quantum portfolio $(a_t ,b_t )$ is given by 

\begin{eqnarray*}
a_t&=&\omega_{0\,1}(t-T,z_t)\nonumber\\
b_t&=&\left(\omega(T-t,z_t)-a_t\,j_t(X)\right)\,e^{-t\,r}\,{\beta_0}^{-1} 
\end{eqnarray*}

\noindent where $z_t$ is defined by $j_t(X)=K\,e^{z_t}$. ( As in the classical case described in Section 1, a \textit{reasonable} price  is defined as one which when invested at time $t=0$ in a mixed portfolio, allows the investor through a self-financing strategy to end up at time $t=T$ with an amount of 

\begin{eqnarray*}
<u\otimes \psi (f),V_T \,u \otimes \psi (f)>&=& <u\otimes \psi (f),( j_T(X) -K)^+\,u \otimes \psi (f)>\\
&=&\max (0, <u\otimes \psi (f), (j_T(X)-K)\,u \otimes \psi (f)>  )
\end{eqnarray*}

\noindent  which is the same as the payoff, had the option been purchased. Here, $u\otimes \psi (f)$ is any vector in the exponential domain of $\mathcal{H} \otimes \Gamma$).

\end{corollary}

\begin{proof} By Theorem 6.1, the reasonable price for a quantum option  is $V_0=F(0,j_0(X))=F(0,X)=u(T,X)=\omega (T,z_0)$. The formulas for $a_t$ and $b_t$ follow from the definition of the portfolio, given in Section 2.
\end{proof}

\end{document}